\documentstyle[12pt]{article}

\makeatletter
\renewcommand\thesubsection{\thesection.\@arabic\c@subsection}
\makeatother

\newcommand{\sect}[1]{\setcounter{equation}{0}\section{#1}}

\newcommand {\beq}{\begin{equation}}
\newcommand {\eeq}{\end{equation}}
\newcommand {\beqa}{\begin{eqnarray}}
\newcommand {\eeqa}{\end{eqnarray}}         
\newcommand {\beqs}{\begin{eqnarray*}}
\newcommand {\eeqs}{\end{eqnarray*}}
\newcommand {\bds}{\begin{displaymath}}
\newcommand {\eds}{\end{displaymath}}
\newcommand {\n}{\nonumber\\}

\newcommand {\bebb}{\begin{thebibliography}}
\newcommand {\eebb}{\end{thebibliography}}      
\newcommand {\bbit}{\bibitem}

\def\pd{\prod}

\def\ra{\rangle}

\def\dg{\dagger}

\def\journal#1&#2(#3){\unskip, \sl #1\ \bf #2 \rm(19#3) }
\def\andjournal#1&#2(#3){\sl #1~\bf #2 \rm (19#3) }

\def\dz{\frac{d}{dz}}

\begin{document}


\begin{flushright}
\end{flushright}

\vskip 1cm

\begin{center}
{\large\bf Solving two-mode squeezed harmonic oscillator and $k$th-order harmonic generation
 in Bargmann-Hilbert spaces}

\vspace{1cm}

{\large Yao-Zhong Zhang}
\vskip.1in

{\em School of Mathematics and Physics,
The University of Queensland, Brisbane, Qld 4072, Australia}

\end{center}

\date{}



\begin{abstract}
We analyze the two-mode squeezed harmonic oscillator and the $k$th-order harmonic generation 
within the framework of Bargmann-Hilbert spaces of entire functions. For the displaced,
single-mode squeezed and two-mode squeezed harmonic oscillators, we derive the exact, 
closed-form expressions for their energies and wave functions.
For the $k$th-order harmonic generation with $k\geq 3$, our result indicates that it does
not have eigenfunctions and is thus ill-defined in the Bargmann-Hilbert space.  
\end{abstract}

\vskip.1in

{\tt 




}  



\setcounter{section}{0}
\setcounter{equation}{0}

\sect{Introduction}

Recently there is renewed interest in formulating and solving dynamical systems involving harmonic modes
in the framework of Bargmann-Hilbert spaces \cite{Schweber67,Reik82,Kus85}. For example,
in \cite{Lee10,Lee11}, we applied the Bargmann-Hilbert space technique to obtain 
the exact solutions of families of quantum non-linear optical as well as spin-boson models. In
 \cite{Braak11,Moroz12,Maciejewski12,Travenec12,Moroz13}, the authors applied the technique to
the quantum  Rabi model, a simple system describing the interaction of a two-level atom with
a harmonic mode. A Bargmann-Hilbert space is a Hilbert space of entire functions 
introduced by Bargmann and Segal. It is a vector space
with typical orthonormal basis $\frac{z^n}{\sqrt{n!}}, n=0, 1, 2, \cdots$. 
Elements in the space are entire functions and the space is equipped with a
well-defined Hermitian scalar product,
\beq
(f,h)=\int\,\overline{f(z)}\,h(z)\,d\mu(z)
\eeq
for any two elements $f(z), h(z)$ in the space, where $d\mu(z)=\frac{1}{\pi}e^{-|z|^2}\,dx\,dy$.
In a Bargmann-Hilbert space, the
harmonic creation and annihilation operators $a^\dagger, a$ can be realized as 
$a^\dagger\rightarrow z,~a\rightarrow \frac{d}{dz}$. This realization enables one to convert
the time-independent Schr\"odinger equation of a dynamical system into a differential equation.
Solutions to the differential equation are entire functions.

In this paper, we apply Bargmann-Hilbert spaces to analyze the two-mode
squeezed harmonic oscillator and $k$th order harmonic generation. For the cases of the displaced,
single-mode squeezed and two-mode squeezed harmonic oscillators, we derive the exact, 
closed-form expressions for their energies and wave functions.
For the $k$th-order harmonic generation with $k\geq 3$, our result shows that it does not have 
eigenfunctions which are entire in the Bargmann-Hilbert space. 
The rest of this paper is as follows. In section \ref{two-mode} we exactly solve the two-mode squeezed
harmonic oscillator. In section \ref{k-photon}, we report our results on our investigation of
solvability of the $k$th order harmonic generation. We draw our conclusions in section \ref{summary}.



\sect{Two-mode squeezed harmonic oscillator}\label{two-mode}
The Hamiltonian of the two-mode squeezed harmonic oscillator reads
\beq
H=\omega(a_1^\dagger a_1+a_2^\dagger a_2)+g\,(a_1^\dagger a_2^\dagger+a_1 a_2),
  \label{2-mode-oscillatorH1}
\eeq
where we assume that the boson modes are degenerate with the same frequency $\omega$,
and $g$ is a real constant. In terms of  the operators $K_\pm, K_0$ defined by
\beq
K_+=a_1^\dagger a_2^\dagger,~~~~K_-=a_1 a_2,~~~~K_0=\frac{1}{2}(a_1^\dagger a_1+a_2^\dagger a_2+1),
   \label{2-mode-boson}
\eeq
the Hamiltonian (\ref{2-mode-oscillatorH1}) can be written as
\beq
H=2\omega\left(K_0-\frac{1}{2}\right)+g\,(K_++K_-).\label{2-mode-oscillatorH2}
\eeq
The operators $K_\pm, K_0$ form the $su(1,1)$ Lie algebra. 
The quadratic Casimir of the algebra, $C=K_+K_--K_0(K_0-1)$, has eigenvalue $\kappa(1-\kappa)$
in the infinite-dimensional unitary irreducible representation of $su(1,1)$ 
known as the positive discrete series ${\cal D}^+(\kappa)$. The parameter $\kappa>0$ is the
so-called Bargmann index. For the two-mode boson realization (\ref{2-mode-boson}),
$\kappa$ can take any positive integers or half integers, i.e. $\kappa=1/2, 1, 3/2,\cdots$. 
Thus the Fock-Hilbert space decomposes into the direct sum of infinite subspaces 
${\cal H}^\kappa$ labeled by $\kappa=1/2, 1, 3/2, \cdots$.

The basis state in the subspace ${\cal H}^\kappa$, denoted as $|\kappa,n\ra,~n=0,1,2,\cdots$, 
has the form
\beq
|\kappa,n\ra=\frac{(a_1^\dagger)^{n+2\kappa-1} (a_2^\dagger)^n}{\sqrt{(n+2\kappa-1)!n!}}|0\ra,
\eeq
and the action of $K_\pm, K_0$ in this representation is given by
\beqa
K_0|\kappa,n\ra&=&(n+\kappa)|\kappa,n\ra,\n
K_+|\kappa,n\ra&=&\sqrt{(n+2\kappa)(n+1)}|\kappa,n+1\ra,\n
K_-|\kappa,n\ra&=&\sqrt{(n+2\kappa-1)n}|\kappa,n-1\ra.\label{2-mode-rep}
\eeqa
Using the Fock-Bargmann correspondence
\beq
a^{\dg} \longrightarrow z , \hspace{1cm} a \longrightarrow \frac{d}{dz},
\hspace{1cm} |0\ra \longrightarrow 1,\label{Fock-Bargmann-correspondence}
\eeq
we can show that the infinite set of monomials
\beq
\Psi_{\kappa,n}(z)=\frac{z^n}{\sqrt{(n+2\kappa-1)!n!}},~~~~n=0,1,2,\cdots,
\eeq
form the basis in the Bargmann-Hilbert subspace associated with the representation (\ref{2-mode-rep}). 
Thus the operators $K_\pm, K_0$ (\ref{2-mode-boson}) have the single-variable differential realization
in the subspace labeled by the Bargmann index $\kappa$,
\beq
K_0=z\frac{d}{dz}+\kappa,~~~~K_+=z,~~~~K_-=z\frac{d^2}{dz^2}+2\kappa\frac{d}{dz},~~~~
  \kappa=1/2, 1, 3/2,\cdots.    \label{su11-diff-rep-2mode}
\eeq
By means of this differential representation (\ref{su11-diff-rep-2mode}), we can express the Hamiltonian 
(\ref{2-mode-oscillatorH2}) (i.e. (\ref{2-mode-oscillatorH1})) as the 2nd-order differential operator
in each Bargmann-Hilbert subspace labeled by $\kappa$,
\beq
H=2\omega\left(z\frac{d}{dz}+\kappa-\frac{1}{2}\right)+g\left(z+z\frac{d^2}{dz^2}
   +2\kappa\frac{d}{dz}\right).   \label{2-mode-diff}
\eeq
Then the time-independent Schr\"odinger equation gives the differential equation for 
wave function $\psi(z)$,
\beqa
&&gz\frac{d^2}{dz^2}\psi(z)+2(\omega z+g\kappa)\frac{d}{dz}\psi(z)
   +\left[gz+2\omega\left(\kappa-\frac{1}{2}\right)-E\right]\psi(z)=0.
\eeqa
With the substitution
\beq
\psi(z)=e^{-\frac{\omega}{g}  (1-\Lambda) z}\varphi(z),
 ~~~~~\Lambda=\sqrt{1-\frac{g^2}{\omega^2}},
\eeq
where $\left|\frac{g}{\omega}\right|<1$, it follows,
\beqa
&&{\cal L}\varphi\equiv \left\{gz\frac{d^2}{dz^2}+2[\omega\Lambda z+g\kappa]\frac{d}{dz}
  +2\kappa\omega\Lambda-\omega-E\right\}\varphi=0.\label{2-mode-diff1}
\eeqa
This differential equation is exactly solvable.  This is seen as follows. First of all,
let us recall the characterization of exact solvability of a differential operator.
A linear differential operator ${\cal L}$ is exactly solvable if it preserves an infinite flag of 
finite-dimensional functional spaces, 
$$
{\cal V}_1\subset {\cal V}_2\subset\cdots \subset {\cal V}_{\cal M}\subset\cdots,
$$
whose bases admit explicit analytic forms, that is there exists a sequence of finite-dimensional 
invariant subspaces ${\cal V}_{\cal M},~{\cal M}=1,2,3,\cdots$,
$$
{\cal L}{\cal V}_{\cal M}\subset {\cal V}_{\cal M},~~~{\rm dim}{\cal V}_{\cal M}<\infty, ~~~
{\cal V}_{\cal M}={\rm span}\{\xi_1,\cdots,\xi_{{\rm dim}{\cal V}_{\cal M}}\}.
$$
In our case, we have, for any positive integer $n$,
\beq
{\cal L}z^n=\left[(2n+2\kappa)\omega\Lambda-\omega-E\right]z^n+n(n+2\kappa-1)gz^{n-1}.
\eeq
It follows that ${\cal L}$ preserves an infinite flag of finite dimensional spaces
${\cal V}_1\subset {\cal V}_2\subset\cdots \subset {\cal V}_{\cal M}\subset\cdots$, 
with explicitly determined subspaces ${\cal V}_{\cal M}=\{1, z, z^2, \cdots, z^{\cal M}\}$, 
and exact solutions are polynomials in $z$ in the Bargmann-Hilbert space. 
We thus seek solutions of the form to the differential equation (\ref{2-mode-diff1}), 
\beqa
\varphi(z)&=&\pd_{i=1}^{\cal M}(z-z_i),~~~~{\cal M}=0,1,2,\cdots,
\eeqa
where $\varphi(z)\equiv 1$ for ${\cal M}=0$, ${\cal M}$ is the degree of the polynomial 
and $z_i$ are roots of the polynomial to be determined. 
Substituting into (\ref{2-mode-diff1}) and dividing both sides by $\varphi(z)$ give rise to 
\beqa
E+\omega-2\kappa\omega\Lambda&=&gz\sum_{i=1}^{\cal M}\frac{1}{z-z_i}
  \sum_{j\neq i}^{\cal M}\frac{2}{z_i-z_j}
  +2[\omega\Lambda z+g\kappa]\sum_{i=1}^{\cal M}\frac{1}{z-z_i}\n
&=&2n\omega\Lambda+\sum_{i=1}^{\cal M}\;\frac{{\rm Res}_{z=z_i}}{z-z_i},\label{evaluating-E}
\eeqa
where ${\rm Res}_{z=z_i}$ are the residues of the right hand side of the first equality
 at the simple poles $z=z_i$, i.e.
\beq
{\rm Res}_{z=z_i}=gz_i\sum_{j\neq i}^{\cal M}\frac{2}{z_i-z_j}+2\omega\Lambda z_i
  +2g\kappa.\label{2-mode-residues}
\eeq
The left hand side (\ref{evaluating-E}) is a constant and the right hand side is a meromorphic function 
with simple poles at $z=z_i$. The right hand side is a constant 
if and only if the coefficient of all the residues at the simple poles are vanishing.  
We thus obtain the energies
\beq
E=-\omega+\left[2{\cal M}+2\left(\kappa-\frac{1}{2}\right)+1\right]\omega\Lambda,\label{2-mode-energy}
\eeq
and the system of algebraic equations satisfied by the roots $z_i$,
\beq
\sum_{j\neq i}^{\cal M}\frac{1}{z_i-z_j}+\frac{\omega}{g}\Lambda+\frac{\kappa}{z_i}=0,~~~~~~i=1,2,\cdots,{\cal M}.\label{2-mode-BEs}
\eeq
The corresponding wave functions are given by 
\beq
\psi(z)=e^{-\frac{\omega}{g}  (1-\Lambda) z}\prod_{i=1}^{\cal M}(z-z_i). \label{2-mode-wavefunction}
\eeq

As examples, we list the first three eigenstates. For ${\cal M}=0$, we have 
$\psi(z)=e^{-\frac{\omega}{g}  (1-\Lambda) z}$. For ${\cal M}=1$, we obtain from (\ref{2-mode-BEs})
the root $z_1=-\frac{\kappa g}{\omega\Lambda}$ and from (\ref{2-mode-wavefunction})
the corresponding wave function  $\psi(z)=e^{-\frac{\omega}{g}  (1-\Lambda) z}
\left(z+\frac{\kappa g}{\omega\Lambda}\right)$. For ${\cal M}=2$, the roots $z_1, z_2$ satisfy
the system of algebraic equations
\beq
\frac{1}{z_1-z_2}+\frac{\omega\Lambda}{g}+\frac{\kappa}{z_1}=0,~~~~~~
\frac{1}{z_2-z_1}+\frac{\omega\Lambda}{g}+\frac{\kappa}{z_2}=0.
\eeq
Solving the two equations simultaneously gives 
\beq
z_1=\frac{-(1+2\kappa)+\sqrt{1+2\kappa}}{2\omega\Lambda}\,g,~~~~~~
z_2=\frac{-(1+2\kappa)-\sqrt{1+2\kappa}}{2\omega\Lambda}\,g.
\eeq
The corresponding wave function is given by
\beq
\psi(z)=e^{-\frac{\omega}{g}  (1-\Lambda) z}\left[z^2+\frac{(1+2\kappa)g}{\omega\Lambda}\,z
  +\frac{\kappa(1+2\kappa)g^2}{2\omega^2\Lambda^2}\right].
\eeq

\sect{$k$th-order harmonic generation}\label{k-photon}
The Hamiltonian of the $k$th-order harmonic generation reads 
\begin{equation}
H=\omega a^\dagger a+g\left[(a^\dagger)^k+a^k\right],\label{k-photon-oscillatorH1}
\end{equation}
where $k=1,2,\cdots$ is any positive integer, and $g$ is a real constant. 
The $k=1$ and $k=2$  cases of  (\ref{k-photon-oscillatorH1}) give 
the Hamiltonians of the displaced and single-mode squeezed harmonic oscillators, respectively
These two oscillator models can be solved by the single-mode Bogoliubov transformation \cite{Emary02}.
For $k\geq 3$, (\ref{k-photon-oscillatorH1}) gives models with higher order harmonic generation.

Introduce the operators $Q_\pm, Q_0$ in terms of the harmonic mode,
\beq
Q_+=\frac{1}{(\sqrt{k})^k}(a^{\dg})^k ,~~~~~
Q_-=\frac{1}{(\sqrt{k})^k}a^k , ~~~~~
Q_0 =\frac{1}{k} \left(a^{\dg}a+ \frac{1}{k} \right).\label{su11-poly-boson}
\eeq
Then in terms of $Q_\pm, Q_0$, the Hamiltonian (\ref{k-photon-oscillatorH1}) can be written as
\beq
H=k\omega\left({Q}_0 -\frac{1}{k^2}\right)+ g\sqrt{k^k}\,\left( {Q}_+ + {Q}_- \right).\label{k-photon-oscillatorH2}
\eeq
It can be shown \cite{Lee10}  that the operators $Q_\pm, Q_0$ form a polynomial algebra of degree $k-1$,
 defined by the commutation relations 
\beqa
[ Q_0,Q_{\pm} ] &=& \pm Q_{\pm}, \n
\left[ Q_+,Q_{-} \right]  &=& \phi^{(k)}(Q_0)-\phi^{(k)}(Q_0-1), \label{su11-poly-alg}
\eeqa
where
\beq
\phi^{(k)}(Q_0)= - \pd_{i=1}^{k} \left( Q_0+\frac{i}{k}-\frac{1}{k^2} \right)
+\pd_{i=1}^k \left(\frac{i-k}{k} - \frac{1}{k^2} \right)
\eeq
is a $k^{th}$-order polynomial in $Q_0$. The Casimir operator of the algebra is given by
\beq
C=Q_-Q_+ +\phi^{(k)}(Q_0)=Q_+Q_-+\phi^{(k)}(Q_0-1).\label{casimir}
\eeq
For $k=1$ and $k=2$, (\ref{su11-poly-alg}) reduces to the Heisenberg and $su(1,1)$ algebras,
 respectively. Thus, the algebra (\ref{su11-poly-alg}) can be viewed as polynomial deformation 
of $su(1,1)$ and Heisenberg Lie algebras.

The realization (\ref{su11-poly-boson}) provides a unitary irreducible representation of the polynomial
algebra, which for $k=2$ reduces to the well-known positive discrete series of $su(1,1)$.
In the realization, the Casimir (\ref{casimir}) takes the particular value,
\beq
C=\pd_{i=1}^k \left(\frac{i-k}{k} - \frac{1}{k^2} \right).\label{casimir-value}
\eeq
If we use $q$ to denote the Bargmann index which labels the basis states of this representation 
in the Fock-Hilbert space ${\cal H}_b $, then it can be shown that $q$ takes $k$ values,
\beq
q= \frac{1}{k^2} ,~ \frac{k+1}{k^2},~\frac{2k+1}{k^2} ,\cdots, ~\frac{(k-1)k+1}{k^2}.\label{q-values}
\eeq
For $k=2$, then $C=\frac{3}{16}$ and $q$ equals to $\frac{1}{4}, \frac{3}{4}$, as expected.
Thus the single-mode boson realization (\ref{su11-poly-boson}) corresponds to the infinite-dimensional
unitary representation with particular $ q $ values (\ref{q-values}), and the Fock
space $ {\cal {H}} _b $ decomposes into the direct sum
$ {\cal H}_b = {\cal H}_{b}^{\frac{1}{k^2}}\oplus \cdots\oplus {\cal H}_b^{\frac{(k-1)k+1}{k^2}}$
of $k$ irreducible components ${\cal H}_b^{\frac{1}{k^2}},..., {\cal H}_b^{\frac{(k-1)k+1}{k^2}}$.

The basis state $|q,n \ra,~n=0,1,\cdots,$ in the irreducible representation space ${\cal H}_b^q$
is then given by \cite{Lee10}
\beq
|q,n\ra= \frac{a^{\dg k(n+q-\frac{1}{k^2})}}{\sqrt{ \left[k\left(n+q-\frac{1}{k^2}\right) \right]!}} 
  |0\ra,~~~~n=0,1,2, \cdots .\label{k-photon-fock-state1}
\eeq
The action  $Q_0, Q_{\pm}$ in this representation reads
\beqa
Q_0|q,n\ra &=& (q+n)|q,n\ra, \n
Q_+|q,n\ra &=& \pd_{i=1}^{k}\left( n+q+ \frac{i k-1}{k^2}\right)^\frac{1}{2}|q,n+1\ra, \n
Q_-|q,n\ra &=& \pd_{i=1}^{k}\left( n+q -\frac{(i-1)k+1}{k^2}\right)^\frac{1}{2}|q,n-1\ra. \label{su11-poly-rep}
\eeqa
By using the Fock-Bargmann correspondence (\ref{Fock-Bargmann-correspondence}),
we can make the following association
\beq
|q,n\ra  \longrightarrow \Psi_{q,n}(z)=\frac{z^n}{\sqrt{\left( k(n+q-\frac{1}{k^2})\right)!}}.
\eeq
It can then be shown that in the Bargmann-Hilbert subspace with basis vectors 
$\Psi_{q,n}(z)$, the operators $Q_\pm, Q_0$ (\ref{su11-poly-boson}) are realized by 
single-variable differential operators
\beqa
{Q}_0 &=& z\frac{d}{dz}+ q, \n
{Q}_+ &=& \frac{z}{(\sqrt{k})^{k}}, \n
{Q}_- &=& z^{-1}(\sqrt{k})^{k} \pd_{j=1}^{k}  \left(z\dz+q- \frac{(j-1)k+1}{k^2} \right).\label{su11-poly-diff}
\eeqa
We remark that there is no singularity in the differential operator expression for $Q_-$.
This is because the $z^{-1}$ term disappears in the expansion of the product in $Q_-$ 
thanks to the fact that $\pd_{j=1}^{k}\left(q- \frac{(j-1)k+1}{k^2} \right)\equiv 0$ 
for all the allowed $q$ values. 

Using the differential realization (\ref{su11-poly-diff}) we can equivalently write 
(\ref{k-photon-oscillatorH2}) 
(i.e. (\ref{k-photon-oscillatorH1})) as the $k$th-order single-variable differential operator in
each Bargmann-Hilbert subspace labelled by index $q$,
\beq
H=k\omega\left(z\frac{d}{dz}+q-\frac{1}{k^2}\right)+g
\left[z+k^k\,z^{-1}\pd_{j=1}^k\left( z\frac{d}{dz}+q-\frac{(j-1)k+1}{k^2}\right)\right].
 \label{k-photon-differentialH}
\eeq
Thus the time-independent Schr\"odinger equation for the model yields 
\beq
\left\{gk^k\,z^{-1}\pd_{j=1}^k\left( z\frac{d}{dz}+q-\frac{(j-1)k+1}{k^2}\right)+gz
+k\omega\left(z\frac{d}{dz}+q-\frac{1}{k^2}\right)-E\right\}\psi=0.  \label{k-photon-diff1}
\eeq

This is a $k$-th order differential equation of Fuchs' type. 
Solutions to this equation must be analytic in the whole complex plane if
$E$ belongs to the spectrum of $H$. In other words, we are seeking solution of the form
\beq
\psi(z)=\sum_{n=0}^\infty K_n(E)\, z^n, \label{series-solution}
\eeq
which converges in the entire complex plane, i.e. solution $\psi(z)$  which is entire. 

Substituting (\ref{series-solution}) into (\ref{k-photon-diff1}), 
we obtain the 3-step recurrence relation,
\beqa
&&K_1(E)+A_0\,K_0(E)=0,\n
&&K_{n+1}(E)+A_n\,K_n(E)+B_n\,K_{n-1}(E)=0,~~~~~~n\geq 1,\label{3-step1}
\eeqa
where
\beqa
A_n&=&\frac{\omega\left(n+q-\frac{1}{k^2}-\frac{E}{k\omega}\right)}{g\,k^{k-1}\,\pd_{j=1}^k\left(n+1+q-\frac{(j-1)k+1}{k^2}\right)},\n
B_n&=&\frac{1}{k^k\,\pd_{j=1}^k\left(n+1+q-\frac{(j-1)k+1}{k^2}\right)}.
\eeqa
The coefficients $A_n, B_n$ have the behavior when $n\rightarrow\infty$,
\beq
A_n\sim a\,n^\alpha,~~~~~~B_n\sim b\,n^\beta
\eeq
with
\beq
a=\frac{\omega}{gk^{k-1}},~~~~\alpha=-k+1,~~~~b=\frac{1}{k^k},~~~~\beta=-k.
\eeq
Thus the asymptotic structure of solutions to the $n\geq 1$ part of (\ref{3-step1}) 
depends on the Newton-Puiseux
diagram formed with the points $P_0(0,0), P_1(1,-k+1), P_2(2,-k)$ \cite{Gautschi67}.
Let $\sigma$ be the slope of $\overline{P_0P_1}$ and $\tau$ the slope of $\overline{P_1P_2}$
so that $\sigma=\alpha$ and $\tau=\beta-\alpha$. Applying the Perron-Kreuser theorem
(i.e. Theorem 2.3 of \cite{Gautschi67}), we have

\vskip.1in
\noindent \underline{\bf $k=1$ case: the displaced harmonic oscillator}. $\sigma=0, \tau=-1$, 
that is $\sigma>\tau$. In this case, the truly 3-term part (i.e. the $n\geq 1$ part) of 
the recurrence relation (\ref{3-step1}) has two linearly independent solutions $K_{n,1},
K_{n,2}$ for which, when $n\rightarrow\infty$
\beq
\frac{K_{n+1,1}}{K_{n,1}}\sim -\frac{\omega}{g},~~~~~~
\frac{K_{n+1,2}}{K_{n,2}}\sim   -\frac{g}{\omega}\,n^{-1}
\eeq
This case belongs to the one treated in \cite{Moroz12}.
So $K^{min}_n\equiv K_{n,2}$ is a minimal solution of the truly 3-term part of (\ref{3-step1}) 
with $k=1$. The corresponding 
infinite power series solution is generated by substituting $K^{min}_n$ for the $K_n$'s in 
(\ref{series-solution}) and converges in the whole complex plane, i.e. it is entire.

\vskip.1in
\noindent \underline{\bf $k=2$ case: the squeezed harmonic oscillator}. $\sigma=-1, \tau=-1$, 
and so $\sigma=\tau=\alpha$. The characteristic equation of the $n\geq 1$ part of (\ref{3-step1}), 
$t^2=0$, has two equal solutions $t_1=t_2=0$. Then all solutions of (\ref{3-step1}) behave
similarly as $n\rightarrow \infty$, viz, 
\beq
\lim_{n\rightarrow\infty}\;{\rm sup}\,\left(|K_n|\,n!\right)^{\frac{1}{n}}=0
\eeq
for all non-trivial solutions of the 2nd equation of (\ref{3-step1}) with $k=2$. 
The zero limit means that the $n\geq 1$ part (i.e. the truly
3-term part) of the recurrence (\ref{3-step1}) possesses a minimal solution $K^{min}_n$ and the 
corresponding infinite power series expansion, obtained by substituting $K^{min}_n$ for the $K_n$'s
in (\ref{series-solution}), converges in the whole complex plane, i.e. it is entire.

\vskip.1in
\noindent \underline{\bf $k\geq 3$ case: anharmonic oscillators}
 $\sigma=-k+1, \tau=-1$, and thus point $P_1$ lies 
below the line segment $\overline{P_0P_1}$ in the Newton-Puiseux diagram. Then 
\beq
\lim_{n\rightarrow\infty}\;{\rm sup}\,\left(|K_n|\,(n!)^{\frac{k}{2}}\right)^{\frac{1}{n}}
   =\frac{1}{\sqrt{k^k}}
\eeq
for all non-trivial solutions of the 2nd equation of (\ref{3-step1}). 
This indicates that solutions to the truly 3-term part of the
recurrence (\ref{3-step1}) with $k\geq 3$ are dominant, and the corresponding 
infinite series expansion (\ref{series-solution})
has a finite radius of convergence proportional to $\sqrt{k^k}$. 
It follows that there does not exist solution to (\ref{k-photon-diff1}) with $k\geq 3$ which is entire. 
We thus conclude that the $k$th-order harmonic generation
model with $k\geq 3$ does not have eigenfunctions (and is ill-defined) in the Bargmann-Hilbert space.
This implies that the Hamiltonian (\ref{k-photon-oscillatorH1}) can not be diagonalized for
$k\geq 3$ using the basis states $|q,n \ra$ (\ref{k-photon-fock-state1}) 
in the Hilbert space ${\cal H}_b$ 
because its eigenstates $|\psi\ra$ is not normalizable (due to the fact that the 
corresponding eigenfunction $\psi(z)$ has a finite radius of convergence in the Bargmann-Hilbert space).

The above analysis still holds for the $k$-photon Rabi model with Hamiltonian
\beq
H_{kR}=\Delta\,\sigma_z+\omega a^\dagger a+g\,\sigma_x\left[(a^\dagger)^k+a^k\right],\label{extraH1}
\eeq
where $\sigma_z, \sigma_x$ are Pauli matrices describing two atomic levels. For degenerate atomic levels
$\Delta=0$, (\ref{extraH1}) has the form of (\ref{k-photon-oscillatorH1}):
\beq
H^{(\Delta=0)}_{kR}=\omega a^\dagger a\pm g\left[(a^\dagger)^k+a^k\right],\label{extraH2}
\eeq
where $\pm$ signs correspond to the two eigenvalues of $\sigma_x$. Now the first term $\Delta\,\sigma_z$
in (\ref{extraH1}) is a bounded spin operator which obviously does not affect the analytic property
of eigenfunctions in the Bargmann-Hilbert space. In other words, eigenfunctions of (\ref{extraH1})
and (\ref{extraH2}) share the same analytic properties and have the same radius of convergence.
Thus another physical consequence of our above result is that the $k$-photon Rabi model is also
non-diagonalizable for $k\geq 3$. This is in sharp contrast to the 
$k$-photon Jaynes-Cummings model which can be exactly solved for all $k$ \cite{Lee10,Lee11}.

In what follows, we will focus on the $k=1, 2$ cases.
By the Pincherle theorem, i.e. Theorem 1.1 of \cite{Gautschi67}, the ratios of successive elements of
the minimal solution sequences $K^{min}_n$ for the $k=1, 2$ cases are expressible 
in terms of infinite continued fractions. 
Proceeding in the direction of increasing $n$, we find
\beq
R_{n}=\frac{K_{n+1}^{min}}{K_n^{min}}=-\frac{B_{n+1}}{~A_{n+1}-}\,\frac{B_{n+2}}{~A_{n+2}-}\,
\frac{B_{n+3}}{~A_{n+3}-}\,\cdots,   \label{continued-fraction}
\eeq
which for $n=0$ gives 
\beq
R_{0}=\frac{K_{1}^{min}}{K_0^{min}}=-\frac{B_{1}}{~A_{1}-}\,\frac{B_{2}}{~A_{2}-}\,
\frac{B_{3}}{~A_{3}-}\,\cdots.   \label{continued-fraction1}
\eeq
Note that the ratio $R_0=\frac{K_1^{min}}{K_0^{min}}$ involves $K_0^{min}$, 
although the above continued fraction expression is obtained from the truly 3-term part 
of (\ref{3-step1}), i.e the recurrence (\ref{3-step1}) for $n\geq 1$. 
However, for single-ended sequences such as those appearing in the infinite power series
expansion (\ref{series-solution}), the ratio $R_0=\frac{K_1^{min}}{K_0^{min}}$ of the first two terms of 
a minimal solution is unambiguously fixed by the $n=0$ part (i.e. the first equation) of 
the recurrence (\ref{3-step1}), namely,
\beq
 R_0=-A_0=-\frac{\omega\left(q-\frac{1}{k^2}-\frac{E}{k\omega}\right)}{g\,k^{k-1}\,
\pd_{j=1}^k\left(1+q-\frac{(j-1)k+1}{k^2}\right)}.   \label{continued-fraction2}
\eeq
 In general, the $R_0$ computed from (\ref{continued-fraction1}) is not the same as that from 
(\ref{continued-fraction2}) (i.e. (\ref{continued-fraction1}) and  (\ref{continued-fraction2})
are not both satisfied)  for arbitrary values of recurrence coefficients $A_n$ and $B_n$. 
As a result, general solutions to the recurrence (\ref{3-step1}) are dominant and are usually
generated by simple forward recursion from a given value of $K_0$. Physical meaningful
solutions are those that are entire in the Bargmann-Hilbert spaces. They can be obtained if $E$
can be adjusted so that equations (\ref{continued-fraction1}) and (\ref{continued-fraction2})
are both satisfied. Then the resulting solution sequence $K_n(E)$ will be purely minimal and
the power series expansion (\ref{series-solution}) will converge in the whole complex plane.

Therefore, if we define the function $F(E)=R_0+A_0$ with $R_0$  
given by the continued fraction in (\ref{continued-fraction1}),  then the
 zeros of $F(E)$ correspond to the points in the parameter space where the condition 
(\ref{continued-fraction2}) is satisfied. In other words, $F(E)=0$ yields the eigenvalue equation,
which may be solved for $E$ by standard nonlinear root-search techniques. 
Only for the denumerable infinite values of $E$ which are the roots of $F(E)=0$, do
we get entire solutions of the differential equations.

As a matter of fact, the spectra for the $k=1, 2$ cases can be determined explicitly. 
As will be seen in the next two subsections, the infinite power series in (\ref{series-solution})
actually truncates for these two cases, so that their solutions are given by polynomials in
Bargmann-Hilbert spaces.

\subsection{Displaced harmonic oscillator}
The displaced harmonic oscillator is the $k=1$ special case of the $k$th-order harmonic 
generation. By (\ref{k-photon-diff1}), the time-independent Schr\"odinger 
equation in the Bargmann-Hilbert space of analytic functions reads \cite{Schweber67}
\beq
(\omega z+g)\frac{d\psi}{dz}+(gz-E)\psi=0.
\eeq
With the substitution 
\beq
\psi(z)=e^{-gz/\omega}\phi(z),
\eeq
the above differential equation reduces to
\beq
\left[(\omega z+g)\frac{d}{dz}-\left(E+\frac{g^2}{\omega}\right)\right]\phi(z)=0.
  \label{oscillator-diff}
\eeq
This differential equation is exactly solvable, and exact solutions are polynomial of
the form
\beq
\phi(z)=\prod_{i=1}^{\cal N}(z-z_i),~~~~{\cal N}=0,1,2, \cdots,\label{oscillator-soln}
\eeq
where $\phi(z)\equiv 1$ for ${\cal N}=0$, ${\cal N}$ is the degree of the polynomial
and  $z_i$ are the roots of the polynomial to be determined. 
Following the procedure similar to that in the last section,
we obtain the energies of the system,
\beq
E=\omega\left({\cal N}-\frac{g^2}{\omega^2}\right),
\eeq
and the set of algebraic equations determining the roots $z_i$, 
$\omega z_i+g=0,~i=1,2,\cdots, {\cal N}$. It follows that
$z_i=-\frac{g}{\omega}$ and the solution (\ref{oscillator-soln}) has the form
\beq
\phi(z)=\prod_{i=1}^{\cal N}\left(z+\frac{g}{\omega}\right) =\left(z+\frac{g}{\omega}\right)^{\cal N}.
\eeq
Thus the wave function of the model is given by
\beq
\psi(z)=e^{-\frac{g}{\omega}z}\left(z+\frac{g}{\omega}\right)^{\cal N}.
\eeq
These expressions for the energies and wave function agree with those in \cite{Schweber67} 
by a different approach.

\subsection{Single-mode squeezed harmonic oscillator}
The Hamiltonian of the single-mode squeezed harmonic oscillator is given by \cite{Emary02}
\begin{equation}
H=\omega a^\dagger a+g\,\left[(a^\dagger)^2+a^2\right],\label{2-photon-oscillatorH1}
\end{equation}
which corresponds to the $k=2$ case of the $k$th-order harmonic generation Hamiltonian 
(\ref{k-photon-oscillatorH1}). 
In the Bargmann-Hilbert space, the time-independent Schr\"odinger equation reads 
\beq
4gz\frac{d^2}{dz^2}\psi(z)+(2\omega z+8gq)\frac{d}{dz}\psi(z)+\left[gz+2\omega\left(q-\frac{1}{4}\right)-E\right]\psi(z)=0,
\eeq
where $q=\frac{1}{4}, \frac{3}{4}$ are the Bargmann index of $su(1,1)$. 
This equation is the $k=2$ special case
of the $k$th order differential equation (\ref{k-photon-diff1}). 

With the substitution
\beq
\psi(z)=e^{-\frac{\omega}{4g}(1-\Omega) z}\varphi(z),~~~~~~
  \Omega=\sqrt{1-\frac{4g^2}{\omega^2}},\label{2-photon-substitution}
\eeq
where $\left|\frac{2g}{\omega}\right|<1$, it follows,
\beqa
&&\left\{4gz\frac{d^2}{dz^2}+[2\omega\Omega z+8gq]\frac{d}{dz}
  +2q\omega\Omega-\frac{1}{2}\omega-E\right\}\varphi=0.  \label{2-photon-diff}
\eeqa
This differential equation is exactly solvable, and exact solutions are polynomials in $z$ 
which  automatically are entire functions
in the Bargmann-Hilbert space. We thus seek solution of the form, 
\beqa
\varphi(z)&=&\pd_{i=1}^{\cal M}(z-z_i),~~~~{\cal M}=0,1,2,\cdots,
\eeqa
where $\varphi(z)\equiv 1$ for ${\cal M}=0$, ${\cal M}$ is the degree of the polynomial solution 
and $z_i$ are the roots of the polynomial to be determined. 
Following the procedure similar to that in the last section, we obtain the energy eigenvalues,
\beqa
&&E=-\frac{1}{2}\omega+\left[2{\cal M}
  +2\left(q-\frac{1}{4}\right)+\frac{1}{2}\right]\omega\Omega,\label{2-photon-energy}
\eeqa
and the set of algebraic equations which determine the roots $z_i$,
\beqa
 \sum_{j\neq i}^{\cal M}\frac{2}{z_i-z_j}+\frac{\omega}{2g}\Omega+\frac{2q}{z_i}=0,~~~~~~
   i=1,2,\cdots, {\cal M}.  \label{2-photon-BEs}
\eeqa
The corresponding wave functions are 
\beq
\psi(z)=e^{-\frac{\omega}{4g}(1-\Omega)z}\prod_{i=1}^{\cal M}(z-z_i).\label{2-photon-wavefunction}
\eeq

Some remarks are in order. The spectrum (\ref{2-photon-energy}) coincides with the 
corresponding result in \cite{Emary02}. This is seen by noting (\ref{2-photon-energy})
is the energy in the Bargmann-Hilbert subspaces labeled by $q=1/4, 3/4$. When $q=1/4$,
we have $2{\cal M}+2(q-1/4)=2{\cal M}$ which corresponds to even integer $n$ in
\cite{Emary02}, while when $q=3/4$, we have $2{\cal M}+2(q-1/4)=2{\cal M}+1$ which
corresponds to odd $n$ in that reference. 

As examples, let us list the first three eigenstates. For ${\cal M}=0$, we have 
$\psi(z)=e^{-\frac{\omega}{4g}(1-\Omega)z}$. For ${\cal M}=1$, we obtain from (\ref{2-photon-BEs})
the root $z_1=-\frac{4q g}{\omega\Omega}$ and from (\ref{2-photon-wavefunction})
the corresponding wave function  $\psi(z)=e^{-\frac{\omega}{4g}(1-\Omega)z}
\left(z+\frac{4q g}{\omega\Omega}\right)$. For ${\cal M}=2$, the roots $z_1, z_2$ satisfy
the system of algebraic equations
\beq
\frac{2}{z_1-z_2}+\frac{\omega\Omega}{2g}+\frac{2q}{z_1}=0,~~~~~~
\frac{2}{z_2-z_1}+\frac{\omega\Omega}{2g}+\frac{2q}{z_2}=0.
\eeq
Solving the two equations simultaneously gives 
\beq
z_1=\frac{-(1+2q)+\sqrt{1+2q}}{\omega\Omega}\,2g,~~~~~~
z_2=\frac{-(1+2q)-\sqrt{1+2q}}{\omega\Omega}\,2g.
\eeq
The corresponding wave function is given by
\beq
\psi(z)=e^{-\frac{\omega}{4g}(1-\Omega)z}\left[z^2+\frac{4(1+2q)g}{\omega\Omega}\,z
  +\frac{8q(1+2q)g^2}{\omega^2\Omega^2}\right].
\eeq

\sect{Conclusions}\label{summary}
We have reported our results on solutions of the two-mode squeezed oscillator and $k$th order
harmonic generation models. These have been achieved through application
of algebraizations and Bargmann-Hilbert spaces. 
We have seen that the algebraizations via either $su(1,1)$ Lie algebra 
or its polynomial deformations decompose the Fock-Hilbert spaces of states 
into direct sums of independent subspaces, 
thus partially diagonalizing the Hamiltonians of the models by bringing them into block-diagonal forms. 
The block-diagonal sectors of the Hamiltonians can be realized as differential operators 
in Bargmann-Hilbert spaces. We have investigated the eigenvalues and eigenfunctions of
the Hamiltonians in these sectors by applying the theory of Bargmann-Hilbert spaces.
For the displaced, single-mode squeezed and two-mode squeezed harmonic oscillators, 
we have obtained the exact, closed-form expressions for their energies and wave functions.
For the $k$th-order harmonic generation with $k\geq 3$, we have shown that it does
not have entire eigenfunctions and thus is ill-defined in the Bargmann-Hilbert space. 
We have argued that same conclusion also holds for the $k$-photon Rabi model with $k\geq 3$.
It is not difficult to see that
the $k$-photon Rabi model Hamiltonian (\ref{extraH1}) possess two (one discrete and one
continuous) degrees of freedom and each of its
states in block-diagonal sector can be labeled by two quantum numbers (energy and parity).
Despite of this fact, the $k\geq 3$ case can not be diagonalized due to the lack of normalizable
eigenstates. Thus this case seems to provide a counter-example to the criteria of 
quantum integrability proposed recently by Braak in \cite{Braak11}. 
A thorough investigation of this point is underway and results will be reported elsewhere.

\vskip.3in
\noindent {\large \bf Acknowledgments} 
\vskip.1in
\noindent This work was supported by the Australian Research Council through 
Discovery Projects grant DP110103434.

\bebb{99}

\bbit{Schweber67}
S. Scheweber, Ann. Phys. {\bf 41}, 205 (1967).

\bbit{Reik82}
H.G. Reik, H. Nusser and L.A. Ribeiro, J. Phys. A: Math. gen. {\bf 15}, 3431 (1982).

\bbit{Kus85}
M. Kus, J. Math. Phys. {\bf 26}, 2792 (1985).

\bbit{Lee10}
Y.-H. Lee, W.-L. Yang and Y.-Z. Zhang, J. Phys. A: Math. Theor. {\bf 43}, 185204 (2010); ibid {\bf 43}, 375211 (2010).

\bbit{Lee11}
Y.-H. Lee, J.R. Links and Y.-Z. Zhang, Nonlinearity {\bf 24}, 1975 (2011).

\bbit{Braak11}
D. Braak, Phys. Rev. Lett. {\bf 107}, 100401 (2011).

\bbit{Moroz12}
A. Moroz, Europhys. Lett. {\bf 100}, 60010 (2012); arXiv:1205.3139v2 [quant-ph].

\bbit{Maciejewski12}
A.J. Maciejewski, M. Przybylska and T. Stachowiak, arXiv:1210.1130v1 [math-ph];
arXiv:1211.4639v1 [quant-ph].

\bbit{Travenec12}
I. Trav\'enec, Phys. Rev. A {\bf 85}, 043805 (2012).

\bbit{Moroz13}
A. Moroz, arXiv:1302.2565v1 [quant-ph], Ann. Phys., in press.

\bbit{Emary02}
C. Emary and R.F. Bishop, J. Math. Phys. {\bf 43}, 3916 (2002); J. Phys. A: Math. Gen. {\bf 35}, 8231 (2002).

\bbit{Gautschi67}
W. Gautschi, SIAM Rev. {\bf 9}, 24 (1967).

\eebb

\end{document}